\documentclass[draft]{agujournal2019}
\usepackage{url} 
\usepackage{lineno}
\usepackage[inline]{trackchanges} 
\usepackage{soul}
\usepackage{siunitx}
\usepackage{amsmath,amssymb}
\usepackage{physics}
\usepackage{subcaption}
\usepackage{soul}
\soulregister\ref7
\soulregister\cite7
\soulregister\citeA7

\draftfalse

\journalname{Geophysical Research Letters}

\begin{document}

%
%


\title{Powering the Galilean Satellites with Moon-moon Tides}

%
%




\authors{Hamish C. F. C. Hay, Antony Trinh, Isamu Matsuyama}


\affiliation{}{Lunar and Planetary Laboratory, University of Arizona, Tucson, AZ 85721, United States}




\correspondingauthor{Hamish Hay}{hhay@lpl.arizona.edu}




\begin{keypoints}
\item Thick subsurface oceans can generate resonant tidal waves in response to moon-moon tidal forcing
\item Enhanced crustal and oceanic energy dissipation due to tidal resonances may alter the thermal-orbital evolution of the Jovian system
\item Measuring moon-moon tidal deformation can help constrain the thickness of subsurface oceans
\end{keypoints}

%
%

%
%


\begin{abstract}
There is compelling evidence for subsurface water oceans among the three outer Galilean satellites, and evidence for an internal magma ocean in the innermost moon, Io. Tidal forces from Jupiter periodically deform these bodies, causing heating and deformation that, if measured, can probe their interior structures. In addition to Jupiter-raised tides, each moon also raises tides on the others. We investigate moon-moon tides for the first time in the Galilean moons, and show that they can cause significant heating through the excitation of high-frequency resonant tidal waves in their subsurface oceans. The heating occurs both in the crust and ocean, and can exceed that of other tidal sources and radiogenic decay if the ocean is inviscid enough. The resulting tidal deformation can be used to constrain subsurface ocean thickness. Our understanding of the thermal-orbital evolution and habitability of the Jovian system may be fundamentally altered as a result.
\end{abstract}

\section*{Plain Language Summary}

The three icy Galilean moons, Europa, Ganymede and Callisto, are thought to contain liquid water oceans beneath their surface, while the innermost moon Io may contain an internal ocean of magma. Jupiter's gravity stretches and squeezes these moons as they orbit the gas giant, heating their interiors through friction. It is essential to understand this process, known as tidal heating, given the unique geophysical structure of ocean worlds and their potential for habitability. In addition to Jupiter, each moon also raises tides on the others, a process that is usually neglected as Jupiter's gravitational attraction is many times larger than that due to the adjacent moons. Here we show that these moon-moon tides cannot in fact be neglected when considering tides as an energy source because they can excite these subsurface oceans near their natural frequencies. By modelling subsurface tidal currents, we find that the corresponding resonant response of the ocean manifests itself through the generation of fast flowing tidal waves, which can release significant amounts of heat into the oceans and crusts of Io and Europa. Our understanding of how ocean worlds in compact systems evolve over time may be altered by the existence of moon-moon tidal resonances.

%
%

%


%
%
%
%

\section{Introduction}

Europa, Ganymede, and Callisto, are all thought to contain global liquid water oceans beneath their hard ice exteriors \cite{zimmer2000subsurface, kivelson2002permanent,saur2015search,hartkorn2017induction}, while Io may contain an internal (partially) molten silicate ocean \cite{khurana2011evidence}. 
Previous studies have shown that diurnal tide-raising forces can produce resonant behavior in icy satellite oceans through the excitation of surface rotational-gravity \cite{tyler2011tidal, tyler2014comparative, matsuyama2014tidal, kamata2015tidal, beuthe2015tidal, beuthe2016crustal, matsuyama2018ocean} and internal inertial \cite{rovira-navarro2019do, rekier2019internal} waves. These tidal waves, or modes, contain substantial kinetic energy, and some fraction of that is converted into heat via turbulent dissipation. Resonant excitation of surface modes occurs only when the ocean is a certain thickness because the surface wave speed, $\sqrt{gh_o}$, where $g$ is surface gravity, is controlled by ocean thickness, $h_o$. Self-gravity and an overlying crust also alters the wave speed . For instance, resonant ocean thicknesses in Europa's ocean at the orbital frequency are generally smaller than a few hundred meters \cite{beuthe2016crustal,matsuyama2018ocean}. Magnetic induction measurements suggest the Galilean satellites' oceans are at least \SI{1}{\km} thick , while oceans up to hundreds of kilometers thick are expected from thermal-chemical modeling \cite{hussmann2002thermal, vance2014ganymedes}. Based on radioscience gravity data, Europa's combined water and ice layer is likely between \SIrange{80}{170}{\km} thick \cite{anderson1998europas}. 
The diurnal tidal forcing, primarily due to a moon's orbital eccentricity and axial tilt (obliquity), hereafter referred to as Jupiter-forcing, is therefore unlikely to excite resonances in the present day oceans of these bodies.
This limits the relevance of Jupiter-raised tides in terms of oceanic heating, although turbulent flow driven by forced-libration of the crust can result in enhanced dissipation \cite{lemasquerier2017libration, wilson2018can}. Thicker oceans can only be resonantly excited if forced at higher frequencies ($|q|>1$, Fig. \ref{fig:eur_a}). Yet, all prior work only considers Jupiter's diurnal tide-raising potential because it is larger than other higher-frequency components. In tidal problems, however, it is essential to not only consider the magnitude of forcing, but also how well a body can respond to that forcing, which is inherently frequency-dependent \cite{greenberg2009frequency}. This letter identifies the high-frequency tidal forces in the Jovian system and calculates how the Galilean satellites, including their oceans, respond to those forces.

\section{Methods}


We calculate the response of the oceans and crusts of the Galilean moons due to tidal forcing at and above the diurnal frequency. This includes high-frequency components of Jupiter's tide-raising potential, and most notably, tidal forcing from the adjacent moons. We calculate the depth-independent (barotropic) dynamical response of a subsurface ocean to these tidal forces by solving the Laplace Tidal Equations (LTE) with linear momentum dissipation using a semi-analytical approach \cite{longuet-higgins1968eigenfunctions, matsuyama2018ocean}, which is benchmarked with  numerical simulation \cite{hay2017numerically, hay2019nonlinear}. The overlying viscoelastic crust and ocean are coupled using membrane theory \cite{beuthe2016crustal}. Further details are given below and in the supporting information.

\subsection{Calculating the full tidal potential}

The oceans and solid-regions of the Galilean satellites are forced with the tide-raising potential from Jupiter and the immediately adjacent moons, assuming the satellites are co-planar and synchronously rotating. Jupiter's tide-raising potential was calculated using Kaula's expansion \cite{kaula1964tidal}. The tide-raising potential due to adjacent moons was calculated assuming circular orbits, using the method outlined in \citeA{hay2019tides} and further verified using the TenGSHui Mathematica library \cite{trinh2019modelling}. We also relaxed circular and co-planar orbit assumptions using TenGSHui but found little effect on the moon-moon forcing potential solutions. Importantly, the inner three moons occupy the Laplace resonance so each moon's tidal forcing frequencies are commensurate with each other. This means that the forcing potentials cannot technically be treated separately as the combined effects of Jupiter and the moons may act to enhance or decrease the total forcing at each frequency.

To be compatible with our primary LTE solution method, we expanded the tidal potential $U^T$ into spherical harmonics using the unnormalized real cosine ($C_{\ell m}$) and sine ($S_{\ell m}$) coefficients;

\begin{linenomath*}
\begin{equation}\label{eq:forcing_general}
U^T(R, \theta, \phi)=\frac{1}{2} \sum_{m=0}^{2} \sum_{\ell =m}^{\infty} \left[ C_{\ell m}(t) - i S_{\ell m}(t) \right]Y_{\ell m} +c . c .
\end{equation}
\end{linenomath*}

\noindent where $Y_{\ell m} = P_{\ell m} (\cos \theta) e^{im\phi}$ is the complex spherical harmonic at degree-$\ell$ and order-$m$, $P_{\ell m}(\cos \theta)$ is the unnormalized associated Legendre function, $c.c$ is the complex-conjugate, and $i$ is the imaginary number. For the purpose of tidal forcing at higher frequencies and that due to other moons, these spherical harmonic expansion coefficients must be further expanded into a Fourier series \cite{hay2019tides}:


\begin{linenomath*}
\begin{equation}
\begin{aligned}[t]
C_{\ell m} (t) &= \sum_{q=-\infty}^{\infty} A_{\ell mq} e^{-iq\omega t},
\end{aligned}
\qquad\qquad 
\begin{aligned}[t]
S_{\ell m} (t) &= \sum_{q=-\infty}^{\infty} B_{\ell mq} e^{-iq\omega t}
\end{aligned}\label{eq:sph_coeffs}
\end{equation}
\end{linenomath*}

\noindent where $\omega$ represents the base forcing frequency, and $q$ is an integer representing multiples of this frequency. The complex Fourier components of the cosine spherical harmonic expansion are $A_{\ell mq}$, while $B_{\ell mq}$ corresponds to the sine expansion coefficients. Details on how these are related to the work of \citeA{hay2019tides} are given in the supporting information. Substituting Eq. \ref{eq:sph_coeffs} into Eq. \ref{eq:forcing_general} and rearranging into real and imaginary components gives;

\begin{linenomath*}
\begin{multline}\label{eq:lte_decomp}
U^{T}(R, \theta, \phi)=\frac{1}{2} \sum_{m=0}^{2} \sum_{\ell =m}^{\infty} \sum_{q=-\infty}^{\infty} \Bigl[ \bigl(\Re(A_{\ell m q}) + \Im(B_{\ell mq})\bigr) - i \bigl(\Re(B_{\ell m q}) - \Im(A_{\ell  m q})\bigr)  \Bigr] Y_{\ell m}e^{-iq\omega t} +c . c .
\end{multline}
\end{linenomath*}


%
%


\noindent For coplanar degree-2 moon-moon tides, the non-zero terms computed are $\Re(A_{20q})$, $\Re(A_{22q})$, and $\Im(B_{22q})$, meaning the spherical harmonic expansion coefficients of the forcing potential are purely real \cite{hay2019tides}. We consider only degree-$\ell=2$ tidal forcing in this letter.


\subsection{Solution of the Laplace Tidal Equations}

We calculated the coupled response of a viscoelastic crust and ocean to tidal forces by solving the extended Laplace Tidal Equations on a sphere which describe the conservation of mass and momentum in a thin, spherical, subsurface ocean. The method is based on the pioneering work of \citeA{longuet-higgins1968eigenfunctions} and is nearly identical to previous work \cite{beuthe2016crustal, matsuyama2018ocean}, except that additional summations of the solution over each forcing frequency are made. Modifications made to the method of \citeA{matsuyama2018ocean} are outlined in the supporting information. We treat Io's magma ocean as a low-viscosity fluid, such that it is adequately described by the LTE. Alternative scenarios similar to \citeA{tyler2015tidal} are explored in the supporting information.

Testing of this method was performed using a numerical finite volume code, Ocean Dissipation in Icy Satellites (ODIS) \cite{hay2017numerically, hay2019nonlinear}. Decomposition of the ocean forcing and response into separate frequency components is unnecessary in ODIS as it solves the LTE in the time-domain. Instead, the ocean was forced directly with the moon-moon tidal potential given in Equation 14 from \citeA{hay2019tides}. We found excellent agreement, with less than \SI{5}{\percent} error between both methods of solution. The resonant modes in Figure \ref{fig:eur_a} were calculated from the eigenvalue problem in Lamb's parameter, outlined in section 4.3 of \citeA{beuthe2016crustal}, and using Eq. 123 from \citeA{beuthe2015tidal} for the non-rotating limit.

\subsection{Solid body tides and heating}


The ocean tide is coupled to a viscoelastic crust using a membrane approach in the LTE \cite{beuthe2008thin,beuthe2016crustal}, where we assume an infinitely rigid mantle. The membrane approximation is excellent for most satellites in predicting the mechanical coupling between the ocean and ice shell, but particularly for large bodies, including the Galilean satellites \cite{matsuyama2018ocean}. We compute dissipation within the crust using the deformation of the ocean top surface, $\eta^{top}$, which is decomposed in an analogous fashion to Eq. \ref{eq:lte_decomp} . Following the approach of \citeA{beuthe2016crustal} and using unnormalized spherical harmonics, we find the time- and surface-averaged crustal dissipation rate to be, 

\begin{linenomath*}
\begin{equation}\label{eq:crustal_diss}
\dot{E}_C = \frac{2 \pi \rho_o g h_o^2}{R^2} \sum_{\ell,m} \ell (\ell+1) N_{\ell m} \sum_{q=-\infty}^{\infty} \frac{\operatorname{Im}\left(\Lambda_{\ell q} \right)}{|q\omega|} \left| \Phi_{\ell m q} \right|^2
\end{equation}
\end{linenomath*}

\noindent where $\Lambda_{\ell q}$ is the degree-$\ell$ membrane spring constant at frequency $q$ \cite<>[Eq. 16]{beuthe2016crustal}, $\Phi_{\ell m q}$ is the degree-$\ell$, order-$m$, and frequency-$q$ ocean velocity potential \cite<>[Eq. C.1]{matsuyama2018ocean}, and

\begin{linenomath*}
\begin{eqnarray}
N_{\ell m} = \frac{\ell (\ell + 1)}{2 \ell + 1} \frac{(\ell + m)!}{(\ell - m)!}
\end{eqnarray}
\end{linenomath*}

\noindent \cite<>[Eq. C.12]{matsuyama2018ocean}. As we assume the mantle is rigid, Eq. \ref{eq:crustal_diss} is identical to Eq. 78 in \citeA{beuthe2016crustal}, with the addition that we sum over each forcing frequency and average over each forcing period, and we use unnormalized spherical harmonics. The membrane constant depends only on the properties of the crust: its thickness, shear modulus, viscosity, and Poisson's ratio (Table S.1). For ice, we assume the crustal viscosity depends on temperature following an Arrhenius relation, and the crustal thermal conductivity is inversely dependent on temperature. We adopt a high melting point viscosity of $10^{17}$ Pa s in this letter. While this value falls outside of the range typically used in tidal and convection studies \cite{barr2009heat}, it is not unreasonable given the uncertainty in the microphysical properties of ice shells (e.g., grain size, impurities, bottom temperature). For rock, the viscosity is assumed to be depth-independent. See Appendix B of \citeA{beuthe2016crustal} and the supporting information for further details.

Dissipation in the core/mantle of these bodies is neglected as we assume the core/mantle is infinitely rigid, considerably simplifying our method. This approximation is valid as tidal dissipation within the core/mantle of these bodies is generally small unless the interior is ultra-low viscosity \cite{segatz1988tidal, bierson2016test}, which can be hard to justify .

\subsection{Equilibrium heat flux} 

We calculated the heat flux required to maintain our assumed crusts in equilibrium in order to provide a reference value for our tidal heating calculations. The crust will remain in thermal equilibrium if the energy leaving the crust is equal to the energy entering at the base, $Q_b = Q_s$. In this scenario, the thickness of the shell, $h_s$, remains constant. Given a fixed surface and bottom temperature, the globally averaged heat flux at the top of a purely conductive crust needed to satisfy this equilibrium for ice and silicate rock is,
\begin{linenomath*}
	\begin{equation}
	F_{ice} =  \frac{a}{h_s} \frac{R-h_s}{R} \ln \left(\frac{T_b}{T_s} \right) \; \; \; \; \; \; F_{rock} =  k \frac{R- h_s}{R} \frac{T_b - T_s}{h_s} \, ,
	\end{equation} 
\end{linenomath*}


\noindent where we set $T_s = \SI{100}{\kelvin}$ and $T_b = \SI{273}{\kelvin}$ for ice \cite{hussmann2002thermal, beuthe2019enceladus}. For Io's silicate crust, $T_b = \SI{1800}{\kelvin}$ \cite{moore2003tidal}. We used $k=\SI{3}{\watt\per\metre\per\kelvin}$ for silicate material, and assumed an $a/T$ dependence for $k$ in ice, where $a=\SI{567}{\watt\per\metre}$ \cite{klinger1980influence}. Only conductive shells are considered in this work for simplicity. Note that we neglect tidal heat production in the crust, so $F$ merely represents a useful reference value to compare our tidal heating estimates against.

\begin{figure}[!t]
	\centering
	\begin{subfigure}{1.0\textwidth} 
		\centering
		\includegraphics[width=0.95\linewidth]{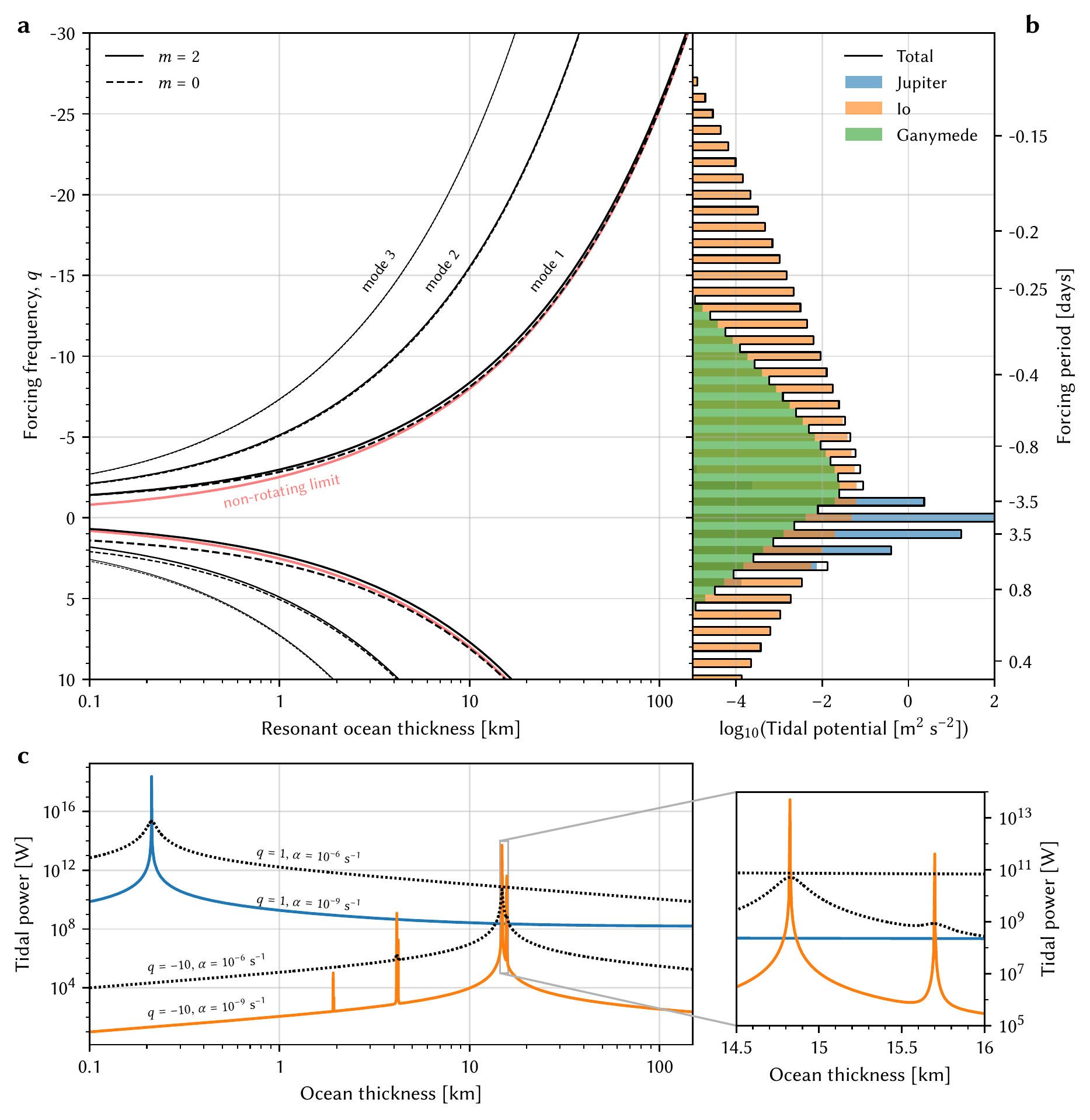} 
		\phantomcaption
		\label{fig:eur_a}
	\end{subfigure}
	\begin{subfigure}{0\textwidth} 
		\phantomcaption
		\label{fig:eur_b}   
	\end{subfigure}
	\begin{subfigure}{0\textwidth} 
		\phantomcaption
		\label{fig:eur_c}   
	\end{subfigure}
	
	\protect\caption{\textbf{Resonant ocean thicknesses at different forcing frequencies.} a) Location of Europa's first three largest resonant rotational-gravity modes as a function of forcing frequency and ocean thickness, for both zonal ($m=0$) and sectoral ($m=2$) degree-2 modes. The sole resonant mode in the non-rotating limit is shown in red.  b) Frequency decomposition of the sectoral degree-2 tidal potential at Europa due to Jupiter, Io, and Ganymede. c) Heating in Europa's ocean and crust due to two frequency components of the tidal forcing; diurnal, eastward Jupiter-forcing (blue) and westward Io-forcing at ten times Europa's orbital frequency (orange), for a drag coefficient $\alpha=\SI{e-9}{\per\second}$. The dotted line uses $\alpha=\SI{e-6}{\per\second}$. The crust is \SI{10}{\km} thick with a melting viscosity of \SI{e17}{\Pa\s}. \label{fig:eur}}
\end{figure} 

\section{Results and Discussion}

\subsection{Moon-moon forcing and resonant ocean thicknesses}

Resonant ocean thicknesses in Europa's ocean are shown in Figure \ref{fig:eur_a} for different multiples, $q$, of its diurnal/orbital frequency, $n_E$, assuming a \SI{10}{\km} thick elastic ice shell. We see that forcing at the orbital frequency ($|q|=1$) cannot excite a resonant response if the ocean is thicker than a few hundred meters. The tide-raising potential due to adjacent moons is smaller in magnitude than that from Jupiter at near-diurnal frequencies but occurs over a much broader frequency spectrum (Fig. \ref{fig:eur_b}). At Europa, Ganymede and in particular Io produce the dominant tidal forcing at all frequencies greater than $\sim 4n_{E}$. Most of the moon-moon tide-raising potential propagates westwards ($q n_E < 0$) across the satellite's surface, while the opposite is true for Jupiter's forcing. See Figure S1 in the supporting information for a time-domain plot of the moon-moon tidal potential.

For demonstration we compare Europa's ocean tidal response at a low- and high-frequency forcing in Figure \ref{fig:eur_c}. A resonance is excited at the eastward-propagating diurnal frequency, $n_E$, which is dominated by Jupiter, if the ocean is $\sim$\SI{200}{\metre} thick. In contrast, the ocean will be resonant if it is $\sim$\SI{14.8}{\km} thick and forced primarily with Io's sectoral tide-raising potential at $q=-10$ times Europa's orbital frequency. The second resonance at $\sim \SI{15.7}{\km}$ is due to Io's zonal tidal forcing. Near the \SI{14.8}{\km} resonant thickness, tidal heating due to Io-forcing surpasses that from Jupiter by several orders of magnitude, assuming a drag coefficient $\alpha \lesssim \SI{e-9}{\per\second}$. This large heating rate, despite the small forcing magnitude, is a result of forcing the ocean at its natural frequency. Kinetic and potential energy imparted by even small tidal perturbations at this frequency will build up in the ocean because the ocean's response propagates with the tidal potential. Energy will continue to be stored until the rate at which kinetic energy is dissipated, parameterized here by $\alpha$, equals the rate at which tidal energy is imparted to the system. This is what limits the maximum possible rate of dissipation in the ocean, at least over short timescales. When $\alpha$ is too large, kinetic energy is dissipated too quickly to allow oceanic perturbations to become sufficiently large, preventing strong resonances from forming. This may be the case for Io if the supposed fluid layer is not a true ``ocean'', but rather only \SIrange{20}{30}{\percent} partial melt, as suggested by \citeA{khurana2011evidence} (Figure S5).  The viscous behaviour of the crust also limits the total amount of heating in resonance (see Figs. S3 and S4 in the supporting information).

\subsection{Linear drag coefficient}
The dashed lines in Figure \ref{fig:eur_c} illustrate that moon-induced tidal heating will not exceed that from Jupiter at the $\sim$\SI{14.8}{\km} resonance if the drag coefficient is larger, \SI{e-6}{\per\second}. Earth's ocean tide studies assume linear drag coefficients $\sim\SI{e-5}{\per\second}$ \cite<e.g.,>[]{egbert2001estimates}. Based on order of magnitude scaling using the more commonly adopted nonlinear drag formalism and using the terrestrial coefficient of bottom drag ($c_D \sim \num{e-3}$), \citeA{matsuyama2018ocean} calculate that the linear drag coefficient in Europa's ocean may be $\sim\SI{e-10}{\per\second}$. Using a similar approach and truncating the LTE to degrees 2 and 3 \cite{chen2014tidal}, we estimate the linear drag coefficient due to non-resonant moon-moon tides to be $< 10^{-12}$ s$^{-1}$ over most frequencies, although this is a minimum estimate (see supporting information).   We accordingly adopt $\alpha=\SI{e-11}{\per\second}$ for the remainder of this letter, although the drag coefficient is ultimately unconstrained. Regardless, Figure 1 shows that over a wide range of drag coefficients,  moon-moon tides cannot be neglected in calculations of oceanic tidal heating. An additional large drag coefficient scenario for Io, corresponding to a ``magma ocean'' contained in a high-porosity asthenosphere, is explored in the supporting information (Figure S5).

\subsection{Heating of the ocean and crust from tidal resonances}

\begin{figure}[!t]
	\centering 
	\includegraphics[width=0.6\linewidth]{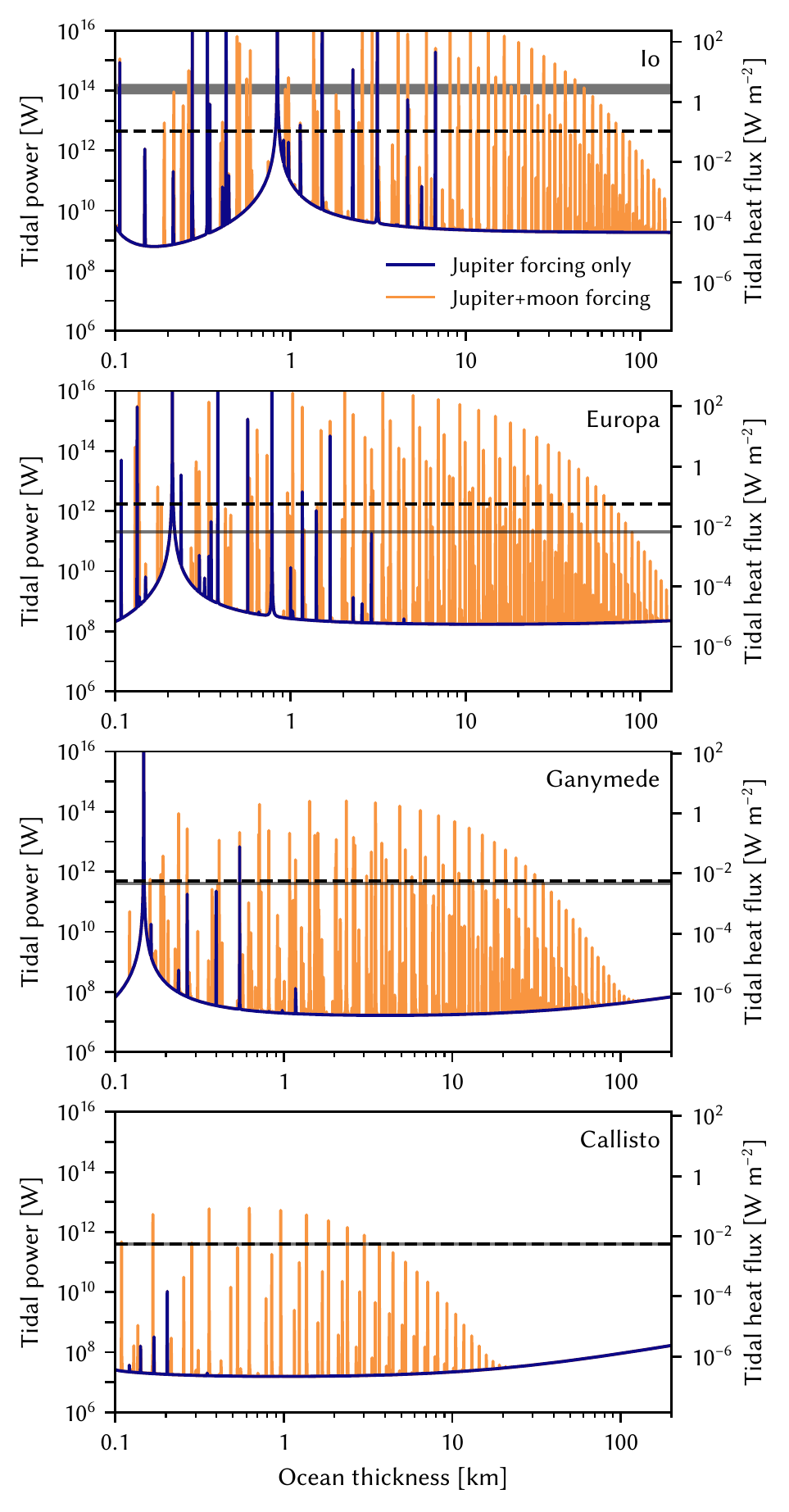} 
	\caption{\textbf{Tidal heating due to Jupiter- and moon-raised tides.} Moon-moon tides are due to the immediate neighboring satellites. Each tidal component includes heating from both the ocean and the solid regions. The ocean's drag coefficient is \SI{e-11}{\per\second}. The grey regions denote estimated radiogenic heating rates \protect\cite{spohn2003oceans}, except for Io where it is the observed heat flux \protect\cite{lainey2009strong}. The dashed lines denote the equilibrium heat flux required to maintain the crustal thickness chosen in our interior structure model.} \label{fig:avg_flux}
\end{figure}

We now calculate tidal heating in the crust and ocean of each Galilean moon as a function of possible ocean thickness using all tidal forcing frequency components (Fig. \ref{fig:avg_flux}). 
We identify several new resonant tidal modes excited by Jupiter alone because we include higher-frequency components of the Jovian tide-raising potential for the first time. These Jupiter-excited resonances can occur in oceans just over \SI{10}{\km} thick for Io, and less than \SI{10}{\km} thick for the other satellites. Many other resonantly excited modes emerge in thicker oceans when including tidal forces from the adjacent moons, because moon-moon tidal forces dominate at high frequencies (Fig. \ref{fig:eur_b}), which thick oceans are most sensitive to (Fig. \ref{fig:eur_a}). These resonant modes can be excited in oceans 10-100 kilometers thick for all moons except Callisto. 
In both Europa and Io, oceans no thicker than $\sim$\SI{70}{\km} may be capable of producing heat in excess of the radiogenic/observed heating rates, and can also conductively maintain crusts of \SIlist{10;50}{\km} thickness, respectively. This is relevant for Europa as a thin crust allows for more rapid transport of organics and oxidants produced at the surface to the ocean beneath \cite{hand2007energy}. The large heating rates due to moon-excited modes may be able to prevent these oceans from freezing past resonant thicknesses, ultimately keeping them tens of kilometers thick.  However, the peak heating rates in Figure \ref{fig:avg_flux} must be taken with some caution. When oceanic flow velocities (and associated heating) become sizeable, saturation may take over or instabilities may form via the nonlinear advection terms neglected in the LTE \cite{lebars2015flows}.  
To order of magnitude, we calculate that this might start to happen above heating rates of $\sim\SI{e12}{\watt}$ on Europa when $\alpha = \SI{e-11}{\per\second}$ (see supporting information). At this point, nonlinearity may act to either locally enhance or suppress tidal currents, making it difficult to predict the effect of nonlinearity on tidal heating.

The crust to ocean heating ratio changes depending on the ocean's drag coefficient and crustal rheology; Europa's tidal heating near resonances is roughly equally partitioned between the ocean and crust for the ice melting viscosity of \protect{\SI{e17}{\Pa\s} used in Figure \ref{fig:avg_flux}}. Further variations in melting viscosity and drag coefficient are explored in Figs. S3 and S4 in the supporting information.
We find low viscosity crusts are more efficient at damping resonances and lowering maximum resonant heating rates, while simultaneously enhancing the non-resonant crustal heating rate. Consequently, if moon-moon tides are to contribute significantly to the tidal heat budget of any of the Galilean moons then high-viscosity (elastic/conductive) crustal behaviour is favoured. This is in contrast to the typical picture of tidal heating in ice shells, where low viscosities, close to the Maxwell critical viscosity, are required to generate significant heat.  

\subsection{Orbital evolution and energy considerations}

The energy dissipated as heat from planet-raised tides on a moon is drawn from the perturbed moon's orbit, decreasing its semimajor axis \cite{kaula1964tidal, murray1999solar} (as observed at Io \cite{lainey2009strong}). In contrast, tidal dissipation within Jupiter transfers energy and angular momentum from its rotation to a moon's orbit, forcing the moon outwards \cite{goldreich1966q, murray1999solar}.
If a moon's ocean enters one of the resonant states described in this letter, removing energy from the orbital system and altering the semimajor axes, then Jupiter would also need to dissipate an equivalent amount of heat to keep the system near equilibrium over geological timescales. 
This balance is not possible with the classical frequency-independent tidal heating model for Jupiter, but may be achieved through tidal excitation of Jupiter's own internal modes \cite{fuller2016resonance}. Even with strong tidal heating in Jupiter replenishing the orbital system's energy, the peak heating rates in Figure \ref{fig:avg_flux} may not be achieved due to the saturation effect of nonlinearity, mentioned above. It is possible, however, that a freezing (thinning) ocean will enter a resonant state only far enough to maintain some equilibrium heating rate (i.e. resonance locking), limiting the impact of nonlinearity.   

The magnitude of moon-moon forcing increases with decreasing separation distance between the two bodies, while the frequencies in the forcing get lower. 
Consequently, if tidal heating causes the conjunction distance between two moons to decrease, moon-moon tidal forces will become larger but resonances in thicker oceans are less likely to be excited due to lower forcing frequencies. If tidal heating has the opposite effect and the conjunction distance increases, moon-moon tidal forces will become weaker \cite{hay2019tides} while the forcing frequencies increase and resonances in thicker oceans are more readily excited. In this way, a non-resonant ocean may become resonant without changing thickness, because the forcing frequencies depend on the conjunction period between the two moons. Small changes in semimajor axis can therefore help modulate tidal heating variations over several orders of magnitude, which could be relevant to periodic volcanism on Io \cite{rathbun2002loki, dekleer2019variability}, if the magma ocean is completely fluid (see supporting information). Additionally, the ocean's natural frequencies will be altered through changes in ocean and shell thickness \cite{kamata2015tidal,beuthe2016crustal,matsuyama2018ocean,hay2019nonlinear}, as well as the ocean's geometry \cite{rovira2020tides}. While resonance locking with an ocean's tidal mode might drive a moon towards equilibrium, orbital changes in the other satellites and the Laplace resonance complicate this picture. Moon-moon tides may have had an effect on the the assembly of the Laplace resonance itself by differential migration, whether by tides raised on Jupiter \cite<e.g.,>[]{yoder1979how}, or by gravitational interactions with a primordial disk \cite<e.g.,>[]{peale2002primordial}. Orbital migration in either scenario will make tidal resonance crossing likely (Figs. S6 and S7 in the supporting information). The combination of heating from tidal resonances in Jupiter and its moons will lead to a complex interplay between the two over solar system history. Moon-moon tides may be able to slow down outward migration due to resonance locking with Jupiter \cite{fuller2016resonance}. 
More work will be needed to understand how the Galilean satellites' interiors and orbits evolve under the influence of moon-excited resonances in their oceans and Jupiter.

\begin{figure}[!t]
	\centering
	
	\begin{subfigure}{\textwidth} 
		\centering
		\includegraphics[width=\textwidth]{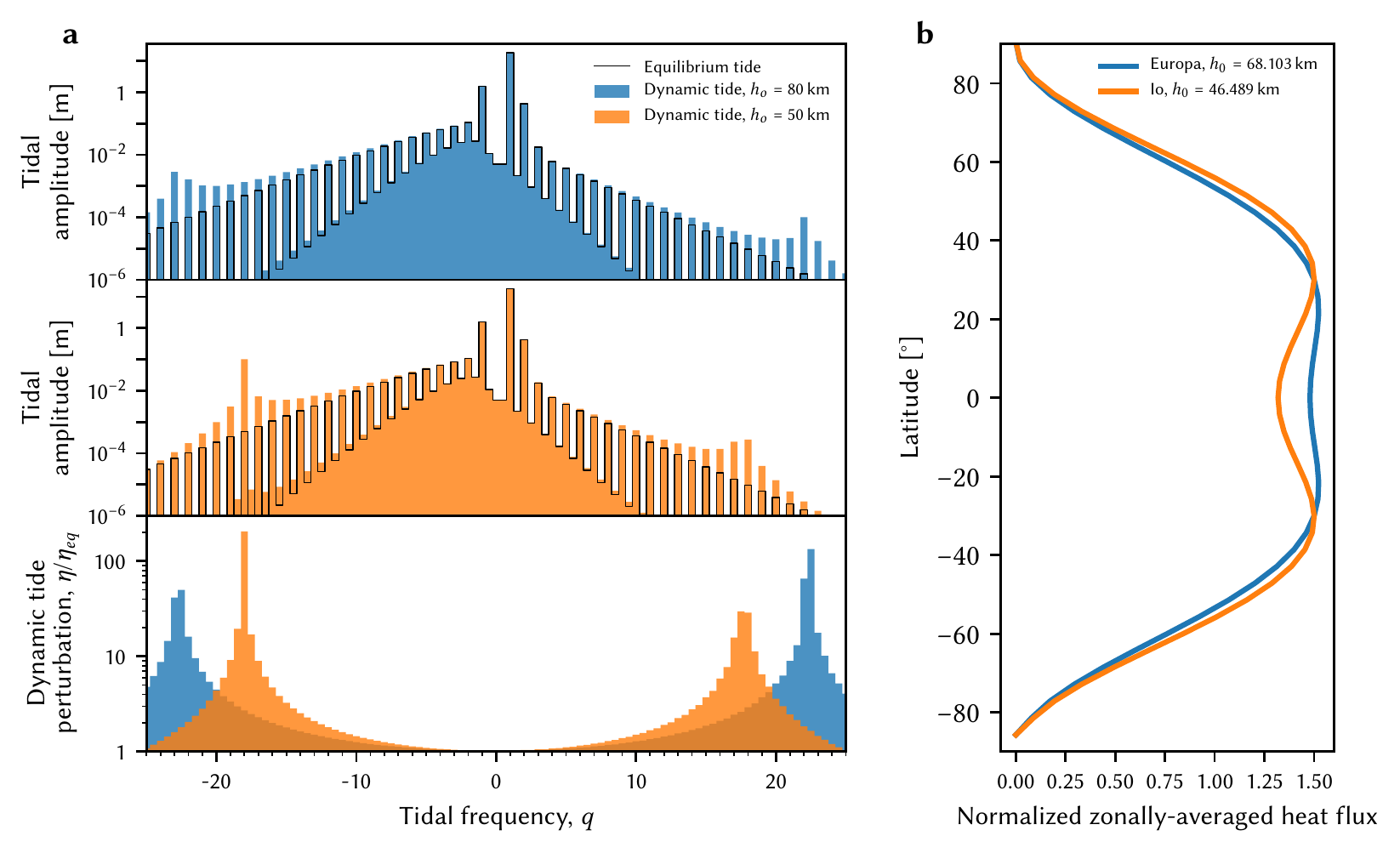} 
		\phantomcaption
		\label{fig:obs_a}
	\end{subfigure}
	\begin{subfigure}{0\textwidth} 
		\phantomcaption
		\label{fig:obs_b}   
	\end{subfigure}
	\vspace{-0.5cm}
	\caption{\textbf{Observable features of moon-moon tides.} a) Frequency spectrum of Europa's surface tidal displacement at the sub-Jovian point for non-resonant \SI{80}{\km} (blue) and \SI{50}{\km} (orange) thick oceans, due to tidal forcing from Io, Ganymede, and Jupiter. The ratio between the dynamical ($\eta$) and equilibrium (black, $\eta_{eq}$) tide displacement is shown in the bottom panel. Negative and positive frequencies correspond to eastward and westward propagating tides, respectively. b) Zonally-averaged tidal heat flux at the ocean's surface for two resonant modes in Io and Europa, using $\alpha=\SI{e-11}{\per\second}$, normalized about the mean heat flux (\SIlist{0.0246; 0.702}{\watt\per\square\metre}, respectively). \label{fig:obs}}	
\end{figure}

\subsection{Observational signatures of moon-moon tides}

The lowest frequency components of the Io-dominated tidal forcing at Europa produce displacements on the order of \SI{10}{\cm} for non-resonant oceans. Dynamical tide effects in thick oceans only become important at higher frequencies. Even for oceans not in resonance, tidal amplitudes become significantly enhanced relative to the commonly assumed equilibrium tide by one to two orders of magnitude  (Fig. \ref{fig:obs_a}). These dynamical tide perturbations, on the order of \SIrange{1}{10}{\cm}, occur at different frequencies depending on the ocean thickness, and are a result of tidal forcing from the adjacent moons. The closer an ocean is to one of its resonant modes, the larger the perturbations become (on the order of $\sim$\SI{1}{\m} up to possibly \SI{1}{\km}, depending on the importance of nonlinearity, see supporting information), but even in non-resonant oceans measurable deviations occur (Fig. \ref{fig:obs_a}). Measuring the frequency-dependent tidal deformation due to moon-moon tides is therefore a unique way to constrain the thickness of a subsurface ocean.

Additional observable signatures may emerge if an ocean is nearly-resonant. The dominant modes due to moon-forcing are westward-propagating tidal waves. These waves produce unique, zonally symmetric patterns of time-averaged heat flux, with heating focused towards low-latitudes and peaking either side of the equator (Fig. \ref{fig:obs_b}). Heightened geological activity at low latitudes would be expected from such a distribution of heat flow, which has been suggested from the locations of chaos terrains on Europa \cite{figueredo2004resurfacing,soderlund2014ocean} and volcanism on Io \cite{veeder2012io, mura2020infrared}, although the polar coverage is poor. The crust would correspondingly be thinner at low latitudes, which could be observable using gravity and topography data. Small-scale turbulent mixing in the ocean may act to diffuse this heating pattern, instead making the time-averaged heat flux more homogeneous than shown in Figure \ref{fig:obs_b}. The maximum possible amount of resonant heating is controlled by the ocean's drag coefficient and crustal viscosity, but the heating pattern in Figure \ref{fig:obs_b} remains unchanged unless $\alpha$ is very large ($> \SI{e-6}{\per\second}$) or the crustal viscosity is very low ($< \SI{e15}{\Pa\s}$). This heating pattern is significantly different to Jupiter-forced tidal heating in the crust, which is enhanced towards the poles \cite{beuthe2013spatial}. Resonant tidal waves are also characterized by fast flows with currents on the order of \SIrange{1}{10}{\metre\per\second}. As this electrically conductive flow interacts with the ambient magnetic field a secondary field will be produced with a strength of around \SI{10}{\nano\tesla} at Europa \cite{tyler2011magnetic}, falling well within the sensitivity range of the magnetometer aboard JUICE \cite{grasset2013jupiter}, and likely Europa Clipper. Dynamical tides produced from resonances may also place torques on the moons, altering their rotation rates \cite{auclair-desrotour2019final}.

\section{Conclusion}

Our study suggests for the first time a mechanism where the ocean could play a crucial role in the heat budget of the Galilean moons, as opposed to previous studies limited to diurnal frequencies where dissipation is often negligible \cite<e.g.,>{chen2014tidal,hay2019nonlinear}. 
In light of this, reexamination of evolution models may be needed in the future. The effect of moon-moon tides may be even larger in the TRAPPIST-1 system if any of the planets contain significant bodies of liquid, as has been suggested \cite{grimm2018nature}. The habitability of closely-packed ocean worlds may depend on these tides.

\acknowledgments
We are grateful to our reviewers Mikael Beuthe and Marc Rovira-Navarro for their thorough and encouraging reviews, Renu Malhotra, William Hubbard, Shane Byrne, Jeffrey Andrews-Hanna and Benjamin Sharkey for thoughtful discussion, and Jessie Brown, Theodore Kareta, and Francis Nimmo for comments on the manuscript. This work was supported by the National Aeronautics and Space Agency (NASA) through the Habitable Worlds program (NNX15AQ88G). All data used in our model are freely available in the literature. Data used to make all figures is in the process of being archived at the University of Arizona Library repository, and is temporarily uploaded as supporting information.

The data to reproduce the figures and all the tidal forcing coefficients are available at \url{https://doi.org/10.5281/zenodo.3902737}.


%
%

\nocite{baland2012obliquity}

\bibliography{master}

%
%
%
%
%

\end{document}